\documentclass{article}
\usepackage{graphicx}

\date{~}

\newcommand{\erf}{\mathop{\rm erf}}

\begin{document}

\title{%
\vskip-6pt \hfill {\rm\normalsize UCLA/00/TEP/33} \\ 
\vskip-6pt \hfill {\rm\normalsize CWRU-P13-00} \\ 
\vskip-6pt \hfill {\rm\normalsize  December 2000} \\
\vskip-9pt~\\
WIMP Annual Modulation with Opposite Phase in Late-Infall Halo Models}
 
\author{Graciela Gelmini\rlap{,}{$^{1}$} Paolo Gondolo\rlap{$^{2}$}
  \\ ~\\
  \small \it ${}^{1}$ Dept.\ of Physics and Astronomy, UCLA (University of
  California, Los Angeles)\\
  \small \it 405 Hilgard Ave., Los Angeles CA 90095, USA\\
  \small \it {\rm gelmini@physics.ucla.edu}
  \\~\\
  \small \it ${}^{2}$ Dept.\ of Physics, Case Western Reserve University
  \\
  \small \it 10900 Euclid Ave., Cleveland, OH 44106-7079, USA\\
  \small \it {\rm pxg26@po.cwru.edu}
  \\~\\
  }
 
\maketitle
 
\begin{abstract} 
  We show that in the late-infall model of our galactic halo by P.~Sikivie the
  expected phase of the annual modulation of a WIMP halo signal in direct
  detection experiments is opposite to the one usually expected. If a
  non-virialized halo component due to the infall of (collisionless) dark
  matter particles cannot be rejected, an annual modulation in a dark matter
  signal should be looked for by experimenters without fixing the phase
  a-priori.  Moreover, WIMP streams coming to Earth from directions above and
  below the galactic plane should be expected, with a characteristic pattern of
  arrival directions.
\end{abstract}

\section{Introduction}

The dark halo of our galaxy may consist of WIMPs (weakly interacting massive
particles).  Direct detection experiments attempt to measure the nuclear recoil
caused by these dark matter particles interacting with the material in a
detector. The most important signature of a halo dark matter signal in these
experiments is its annual modulation~\cite{modulation}. The WIMP interaction
rate, the energy deposited per collision, as well as the annual modulation of
the signal depend strongly on the velocity distribution of the halo dark matter
particles with respect to the detector.  It is therefore very important to
explore the possibility of halo particles velocity distributions which differ
from the one usually assumed. The standard assumption is of a virialized dark
halo, on average at rest in the rest frame of the galaxy, with a gaussian
velocity distribution truncated due to the escape velocity from the galaxy, of
about 600 km/s. The average WIMP velocity on Earth in then due to the motion
of the Earth with respect to the galaxy (at about 200 km/s).  This velocity is
maximal around June 2 each year, when the velocity of the Earth around the Sun
adds up maximally to the velocity of the Sun with respect to the galaxy, and
minimal six months later, around December 2 each year. 

A non-standard halo model is the late-infall model of Refs.~\cite{LImodel},
modified in recent years by P.~Sikivie and collaborators
\cite{SI,STW,Slecture}. It assumes a non-virialized dark halo in which
collisionless dark matter particles falling into the galaxy oscillate in and
out many times. At a given location in the galaxy, multiple flows of particles
are possible, each having a specific velocity in a specific direction. The
non-virialized flows of dark matter particles produce a velocity distribution
completely different from the standard truncated Gaussian. The particular
late-infall model of Sikivie in Ref.~\cite{Slecture} is a self-similar axially
symmetric infall model with net angular momentum and parameters adjusted to
describe well our galaxy. For this model, which we call Sikivie's LI model or
SLI model from now on, the local velocities and densities of the first twenty
pairs of flows are given in Table 1 of Ref.~\cite{Slecture}. We reproduce this
table in our Table 1 for convenience.  The first pair of flows corresponds to
particles coming into the galaxy for the first time from opposite sides of it,
the second to those passing for the second time, etc.

Here we would like to clearly expose the differences that this particular late
infall halo model by Sikivie implies for direct dark matter detection
experiments.

\begin{table}
\begin{center}
\begin{tabular}{|c|c|c|c|c|c|}
\hline
$i$ & $\rho_i$ & $v_{iX}$ & $v_{iY}$ & $v_{iZ}$ & $u_i^+$, $u_i^-$\\
    & ($10^{-26}$ g/cm$^3$) & (km/s) & (km/s) & (km/s) & (km/s) \\
\hline
$~1^\pm$ & 0.4   &        0  &    140 &   $\pm605$ & 605, 619 \\
$~2^\pm$ & 1.0   &        0  &    255 &   $\pm505$ & 499, 513 \\
$~3^\pm$ & 2.0   &        0  &    350 &   $\pm390$ & 401, 414 \\
$~4^\pm$ & 6.3   &        0  &    440 &   $\pm240$ & 312, 322 \\
$~5^\pm$ & 9.2   &  $\pm190$ &    440 &         0  & 274, 288 \\
$~6^\pm$ & 2.9   &  $\pm295$ &    355 &         0  & 310, 329 \\
$~7^\pm$ & 1.9   &  $\pm330$ &    290 &         0  & 325, 345 \\
$~8^\pm$ & 1.4   &  $\pm350$ &    250 &         0  & 340, 360 \\
$~9^\pm$ & 1.1   &  $\pm355$ &    215 &         0  & 346, 365 \\
$10^\pm$ & 1.0   &  $\pm355$ &    190 &         0  & 348, 368 \\
$11^\pm$ & 0.9   &  $\pm355$ &    170 &         0  & 351, 370 \\
$12^\pm$ & 0.8   &  $\pm350$ &    150 &         0  & 350, 369 \\
$13^\pm$ & 0.7   &  $\pm345$ &    135 &         0  & 349, 368 \\
$14^\pm$ & 0.6   &  $\pm340$ &    120 &         0  & 349, 368 \\
$15^\pm$ & 0.6   &  $\pm330$ &    110 &         0  & 343, 362 \\
$16^\pm$ & 0.55  &  $\pm325$ &    100 &         0  & 342, 360 \\
$17^\pm$ & 0.50  &  $\pm320$ &     90 &         0  & 341, 360 \\
$18^\pm$ & 0.50  &  $\pm310$ &     85 &         0  & 335, 353 \\
$19^\pm$ & 0.45  &  $\pm305$ &     80 &         0  & 333, 350 \\
$20^\pm$ & 0.45  &  $\pm300$ &     75 &         0  & 330, 348 \\
\hline
\end{tabular}
\end{center}
\caption{Local densities and velocities of the first 20 pairs of flows in the
  self-similar infall model of ref.~\cite{STW}. The first five columns are from
  Table 1 in Ref.~\cite{Slecture}. The last column gives the flow speeds
  with respect to the Sun (as obtained by us).}
\end{table}

\section{Characteristics of the WIMP Wind}

In the first comparison between the standard halo model and Sikivie's late
infall model, we consider the flux of WIMPs as a function of solid angle for an
observer moving with the Sun. If ${\bf v}$ is the velocity of a WIMP with
respect to the galaxy, its velocity ${\bf u}$ with respect to the Sun is simply
given by a galilean transformation
\begin{equation}
{\bf u} = {\bf v} - {\bf v}_{\odot} ,
\end{equation}
where
\begin{equation}
{\bf v}_{\odot} = {\bf v}_{\rm LSR} + {\bf v}'_{\odot} .
\end{equation}
$ {\bf v}_{\odot} $ is the velocity of the Sun with respect to the galactic
rest frame, ${\bf v}_{\rm LSR}$ is the velocity of the Local Standard of Rest
(LSR), which is in the direction of galactic rotation, and ${\bf v}'_{\odot}$
is the peculiar velocity of the Sun, i.e.\ its velocity with respect to the
LSR. In a coordinate system in which $X$ points toward the galactic center, $Y$
toward the direction of galactic rotation, and $Z$ toward the North Galactic
Pole, we adopt
\begin{equation}
{\bf v}_{\rm LSR} = (0,220,0) \, {\rm km/s},
\end{equation}
for the velocity of the Local Standard of Rest, and
\begin{equation}
{\bf v}_{\odot} = (10,13,7) \, {\rm km/s}
\end{equation}
for the Sun peculiar velocity. (The uncertainty in the Sun peculiar velocity is
of the order of 0.2 km/s in the $Z$ direction and of as much as 3 km/s in the
$X$ and $Y$ directions \cite{solarmotion}; the corresponding uncertainty in the
phase constant of the modulation we discuss below is of several days.)

With these preliminaries, the flux of WIMPs arriving at the Sun from within the
solid angle $d\Omega$ around the direction $\hat{\bf n}$ is
\begin{equation}
\label{eq:flux}
\frac{d\Phi}{d\Omega} = \frac{\rho}{m} \, 
\int u f_{\odot}(-u \hat{\bf n}) \cdot u^2 du .
\end{equation}
Here $\rho$ is the local halo density, $m$ is the WIMP mass, and
$f_{\odot}({\bf u}) d^3 u$ is the fraction of WIMPs with velocities with
respect to the Sun within $d^3 u$ around ${\bf u}$.  The WIMP velocity
distribution in the rest frame of the Sun $f_{\odot}({\bf u})$ is related to
the WIMP velocity distribution in the galactic rest frame $f({\bf v})$ through
\begin{equation}
f_{\odot}({\bf u}) = f({\bf u}+{\bf v}_{\odot}) .
\end{equation}

Below we plot the WIMP flux $d\Phi/d\Omega$ as a function of the arrival
direction $\hat{\bf n}$ in galactic coordinates $(l,b)$, $\hat{\bf n} = (\cos b
\cos l, \cos b \sin l, \sin b)$. We do this for two models of the velocity
distribution: the standard halo model and the late-infall model of Sikivie and
collaborators.

In the standard halo model, the WIMP velocity distribution is assumed to be a
gaussian with velocity dispersion $\overline{v}_0/\sqrt{2}$ truncated at the
escape velocity $v_{\rm esc}$,
\begin{equation}
  f_{\rm std}({\bf v}) = \cases{
    \displaystyle \frac{1}{N_{\rm esc} \pi^{3/2} \overline{v}_0^3} 
    \, e^{-{\bf v}^2\!/\overline{v}_0^2} , 
    & for $ |{\bf v}| < v_{\rm esc} $ \cr
    0 , & otherwise, }
\end{equation}
where $\rho$ is the local WIMP density, $m$ is the WIMP mass, and $ N_{\rm esc}
= \erf(z_0) - 2 z_0 \exp(-z_0^2) / \pi^{1/2} $ with $z_0 = v_{\rm
  esc}/\overline{v}_0$ is a normalization factor.  For the sake of
illustration, we take $\overline{v}_0 = 220$ km/s and $v_{\rm esc} = 650$
km/s. Other values do not change our conclusions.

A simple integration gives the flux per unit solid angle as
\begin{equation}
\frac{d\Phi_{\rm std}}{d\Omega} = \frac{\rho v_{\odot} }{m} \, h_0(l,b) ,
\end{equation}
where $v_{\odot} = |{\bf v}_{\odot}|$ and
\begin{eqnarray}
&&  h_0(l,b) = 
  \frac{e^{-x_0^2}}{4 \pi^{3/2} x_0 N_{\rm esc}} \, \Big\{ 
  2(1+y_0^2) 
  \nonumber \\ && 
  \hspace{1.0in}
  - \sqrt{\pi} e^{y_0^2} y_0 (3+2y_0^2) 
  [\erf(\sqrt{z_0^2+y_0^2-x_0^2}) - \erf(y_0)] 
  \nonumber \\ &&
  \hspace{1.0in}
  - 2 e^{x_0^2-z_0^2} 
  ( 1+4y_0^2+z_0^2-x_0^2-3y_0\sqrt{z_0^2+y_0^2-x_0^2} ) \Big\} .
\end{eqnarray}
with $x_0 = v_{\odot}/\overline{v}_0$, $ y_0 = - {\bf v}_{\odot} \cdot \hat{\bf
  n}/\overline{v}_0$, and $ z_0 = v_{\rm esc}/\overline{v}_0$.

In the late-infall halo model SLI, the WIMPs belonging to stream $i$ have all
the same velocity ${\bf v}_i$ and they contribute a density $\rho_i$ to the
local WIMP density. The values of ${\bf v}_i$ and $\rho_i$ are given in Table
1. The model of Ref.~\cite{Slecture} does not include a velocity dispersion of
the dark matter particles in the flows. In this case, the WIMP velocity
distribution in the SLI model is
\begin{equation}
f_{\rm SLI}({\bf v}) = \frac{1}{\rho} \, \sum_i \rho_i \, 
\delta( {\bf v} - {\bf v}_i ) .
\end{equation}
The corresponding angular distribution of the
WIMP flux simply consists of isolated points,
\begin{equation}
\frac{d \Phi_{\rm SLI}}{d\Omega} = \sum_i \frac{ \rho_i u_i}{m} \,
\delta(\hat{\bf u}_i + \hat{\bf n}) .
\end{equation}
Here ${\bf u}_i = {\bf v}_i - {\bf v}_{\odot}$, $ u_i = | {\bf u}_i | $, and
$\hat{\bf u}_i = {\bf u}_i/u_i $.  If we introduce by hand a velocity
dispersion $\overline{v}_i/\sqrt{2}$ for each flow, and assume a gaussian
distribution of velocities for each flow, the velocity distribution becomes
\begin{equation}
  f_{\overline{\rm SLI}}({\bf v}) = \sum_i
    \frac{\rho_i}{\rho \pi^{3/2} \overline{v}_i^3} 
    \, e^{-{\bf v}^2\!/\overline{v}_i^2} , 
\end{equation}
and the WIMP flux per unit solid angle
\begin{equation}
\frac{d\Phi_{\overline{\rm SLI}}}{d\Omega} = 
\sum_{i} \frac{\rho_i u_i }{m} \, h_i(l,b) ,
\end{equation}
where 
\begin{equation}
h_i(l,b) = \frac{e^{-x_i^2}}{4 \pi^{3/2} x_i} \Big\{ 2(1+y_i^2) - 
\sqrt{\pi} e^{y_i^2} y_i (3+2y_i^2) [1-\erf(y_i)] \Big\} 
\end{equation}
with $x_i = u_i/\overline{v}_i$ and $y_i = {\bf u}_i \cdot
\hat{\bf n}/\overline{v}_i$.  In the examples below, we set
$\overline{v}_i = 30 $~km/s.

\begin{figure}
\begin{center}
\includegraphics[width=\textwidth]{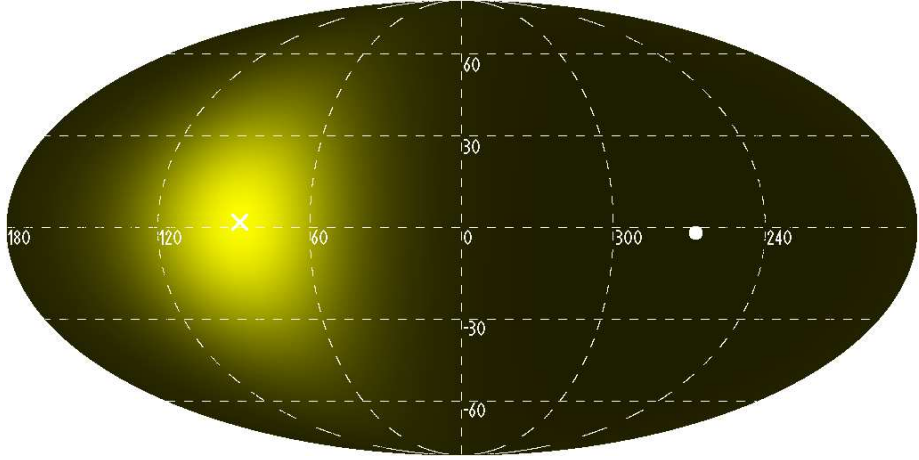}
\end{center}
\caption{
  Sky map in galactic coordinates of the WIMP flux as seen by an observer
  moving with the Sun for the standard halo model with $\overline{v} = 220$
  km/s and $v_{\rm esc}=650$ km/s. Lighter colors indicate larger flux
  intensities.  The WIMP wind comes mostly from the direction of the
  Sun motion (white cross).}
\end{figure}
\begin{figure}
\begin{center}
\includegraphics[width=\textwidth]{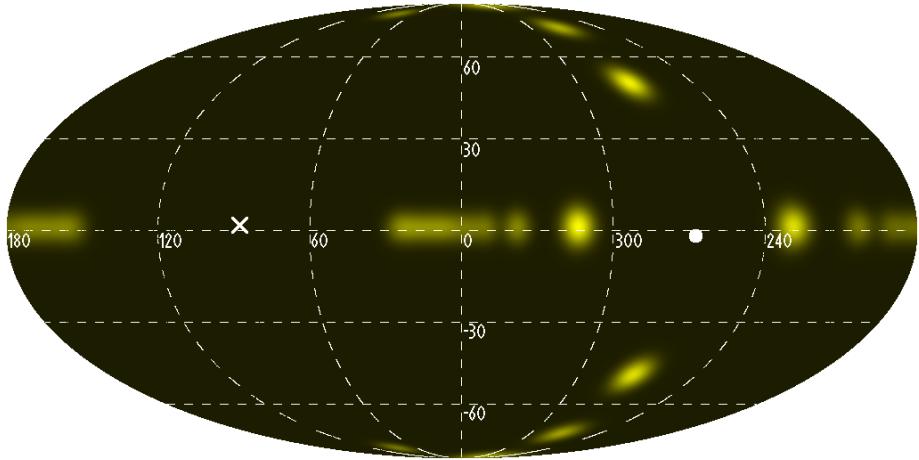}
\end{center}
\caption{
  Sky map in galactic coordinates of the WIMP flux as seen by an observer
  moving with the Sun for Sikivie's late-infall halo model, smoothing each
  stream with a gaussian with velocity dispersion $\overline{v}_i = 30$ km/s.
  Lighter colors indicate larger flux intensities. WIMPs come from the
  direction of the streams (the bright spots). The most intense streams lie
  around the direction opposite to the Sun motion (white circle).  This
  reverses the phase of the annual modulation of a halo WIMP signal due to the
  motion of the Earth around the Sun.}
\end{figure}

Figs.~1 and 2 show sky maps of the flux of dark matter particles in the
standard halo model (Fig.~1) and in the SLI halo model (Fig.~2). These are the
fluxes according to an observer at rest with the solar system.  The sky maps
in Figs.~1 and~2 are equal-area projections of the celestial sphere in galactic
coordinates.\footnote{They are Mollweide projections. The relationship between
  the coordinates $(x,y)$ on the map and the galactic coordinates $(b,l)$ is $
  x = -(2\sqrt{2} l \cos\theta)/\pi $, $y = \sqrt{2} \sin \theta $, where
  $\theta$ is given by $2\theta+\sin(2\theta)=\pi\sin b$.  The inverse formulas
  are $ b=\arcsin\left\{[(2\theta+\sin(2\theta)]/\pi\right\}$, $ l=-(\pi
  x)/(2\sqrt{2} \cos{\theta})$, where $\theta = \arcsin(y/\sqrt{2})$.  }  The
galactic center is at the center, the galactic north (south) pole is at the top
(bottom), the galactic plane is horizontal passing through the center of the
map. The Sun is moving towards the direction indicated by the white cross (Sun
apex). The white circle indicates the direction opposite to the Sun's motion
(Sun anti-apex).  The color in the maps shows the intensity of the WIMP flux
coming from each direction (a lighter color indicates a larger intensity).

In the standard halo model of Fig.~1, most WIMPs come from the direction
towards which the Sun is moving.  In the SLI model of Fig.~2, WIMPs come from
the direction of the streams (the bright spots). In the figure, each stream has
been convolved with a Maxwellian velocity distribution with velocity dispersion
$\overline{v}_i=30$ km/s (otherwise the WIMPs of each stream would come only
from one point in the sky). Fig.~2 clearly shows that in the SLI halo model
most WIMPs come from directions in the hemisphere opposite to the Sun motion.
As a consequence, the average ``WIMP wind'' velocity on Earth, as pointed out
by Sikivie himself in Ref.~\cite{Slecture}, is reversed with respect to that in
the standard halo model. This is because most halo particles in the SLI model
move in the direction of the Sun's motion with a speed larger than that of the
Sun.  Therefore in the SLI model the annual modulation of a galactic WIMP
signal has a phase opposite to the one usually assumed. Notice in Fig.~2 that
in the SLI halo model many WIMPs come to Earth also from directions above and
below the galactic plane. The directions of the streams form a characteristic
``diamond'' or ``quad'' pattern around the Sun anti-apex: the most intense WIMP
fluxes are concentrated on the galactic plane and on the $l=270^\circ$
meridian.

\section{Annual Modulation of WIMP Fluxes} 

The time dependence of the galactic WIMP signal arises from the annual
variation of the Earth velocity with respect to the Sun. Direct detection
experiments lacking directional capabilities are sensitive to the WIMP flux
integrated over the whole sky. This sky-integrated flux is the product of the
local WIMP number density and of the mean speed of the WIMPs with respect to
the Earth. It's the latter that is modulated. So here we study the time
variations of the mean WIMP speed with respect to the Earth. (Notice that this
is the mean speed, not the mean velocity.)

First we write an expression for the velocity of the Earth. We neglect the
ellipticity of the Earth orbit and the non-uniform motion of the Sun in right
ascension (an error of less than 2 days in the position of the modulation
maximum and minimum). Hence we write the velocity of the Earth in terms of the
Sun ecliptic longitude $\lambda(t)$ as
\begin{equation}
{\bf v}_{\oplus}(t) = v_{\oplus} \, 
\left[ \hat{\bf e}_1 \sin \lambda(t) - \hat{\bf e}_2 \cos \lambda(t) \right] ,
\end{equation}
where $v_{\oplus} = 2 \pi {\rm A.U.}/{\rm yr} = 29.8$~km/s is the orbital speed
of the Earth, and the unit vectors $\hat{\bf e}_1$ and $\hat{\bf e}_2$ are in
the direction of the Sun at the spring equinox and at the summer solstice,
respectively. In galactic coordinates,
\begin{eqnarray}
&&  \hat{\bf e}_1 = ( -0.0670, 0.4927, -0.8676 ) , \\
&&  \hat{\bf e}_2 = ( -0.9931, -0.1170, 0.01032 ) . 
\end{eqnarray}
The Sun ecliptic longitude $\lambda(t)$ can be expressed as a function of 
time $t$ in years with $t=0$ at January 1 as
\begin{equation}
\lambda(t) = 360^\circ \,  (t - 0.218) .
\end{equation}
Here 0.218 is the fraction of year before the spring equinox (March 21).

We give now the mean WIMP speed with respect to the Earth. It is 
\begin{equation}
  \langle v(t) \rangle = \int u \, f_{\oplus}({\bf u},t) \, d^3 u ,
\end{equation}
where $f_{\oplus}({\bf u},t)$ is the WIMP velocity distribution in the rest
frame of the Earth. In terms of the velocity distribution in the galactic rest
frame $f({\bf v})$, 
\begin{equation}
  f_{\oplus}({\bf u},t) = f( {\bf u} + {\bf v}_{\odot} + {\bf v}_{\oplus}(t) )
  .
\end{equation}
So we can also write
\begin{equation}
  \langle v(t) \rangle = 
  \int \big| {\bf v} - {\bf v}_{\odot} - {\bf
    v}_{\oplus}(t) \big|  \, f({\bf v}) \, d^3 v .
\end{equation}

In the standard halo model, the mean WIMP speed on Earth results
\begin{equation}
\langle v_{\rm std}(t) \rangle = v_0(t) \, g_0(t) ,
\end{equation}
where $v_0(t) = \big| {\bf v}_{\odot} + {\bf v}_{\oplus}(t) \big| $ and
\begin{equation}
g_0(t) = \frac{1}{N_{\rm esc}} \left[
\frac{e^{-x_0^2}}{x_0\sqrt{\pi}} + \left( 1+\frac{1}{2x_0^2}
\right) \erf(x_0) - 
\sqrt{\frac{2}{\pi}} \frac{1+z_0^2+x_0^2/3}{x_0} e^{-z_0^2} \right] 
\end{equation}
with $x_0 = v_0(t)/\overline{v}_0$ and $z_0 = v_{\rm esc}/\overline{v}_0$.

In the SLI halo model with no velocity dispersion,
\begin{equation}
\langle v_{\rm SLI}(t) \rangle = \sum_{i} 
\frac{\rho_i v_i(t)}{\rho} 
\end{equation}
with $v_i(t) = | {\bf v}_i - {\bf v}_{\odot} - {\bf v}_{\oplus}(t) |$.
Including a velocity dispersion as described above,
\begin{equation}
\langle v_{\overline{\rm SLI}}(t)  \rangle = \sum_{i} 
\frac{\rho_i v_i(t)}{\rho} \, g_i(t) ,
\end{equation}
where
\begin{equation}
  g_i(t) = \frac{e^{-x_i^2}}{x_i\sqrt{\pi}} + \left( 1+\frac{1}{2x_i^2}
  \right) \erf(x_i)
\end{equation}
with $x_i = | {\bf v}_i - {\bf v}_{\odot} - {\bf v}_{\oplus}(t)
|/\overline{v}_i$.

We have plotted the mean WIMP speed $\langle v(t) \rangle$ as a function of
time in Fig.~3 for the standard halo model and the SLI halo model with and
without a velocity dispersion.  The time axis starts January 1 and covers a
year.

\begin{figure}
\begin{center}
\includegraphics[width=\textwidth]{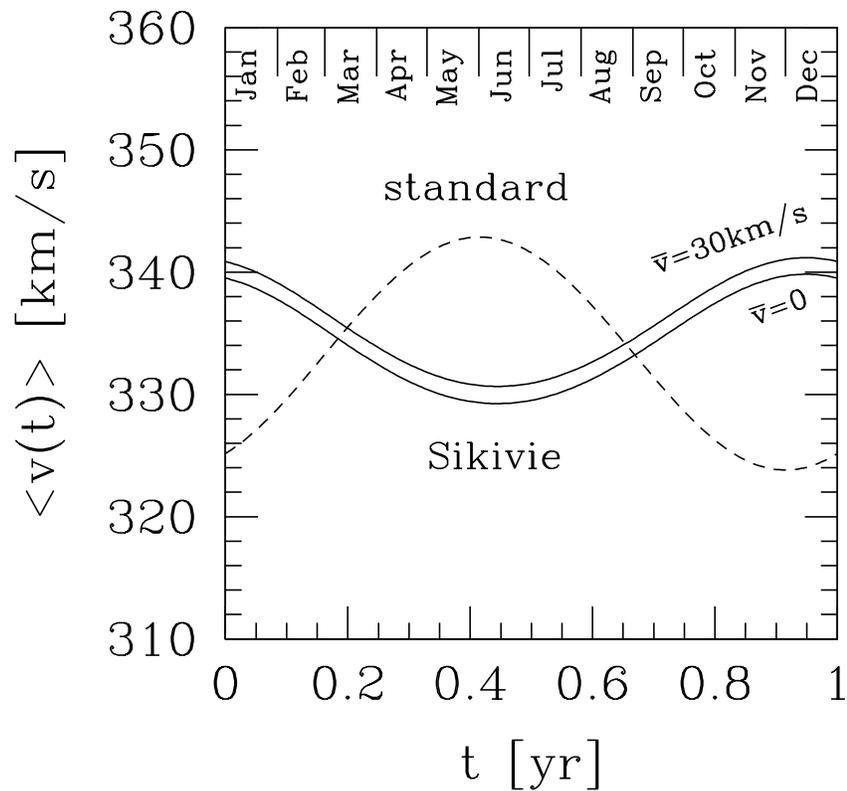}
\end{center}
\caption{Annual modulation in the mean WIMP speed on Earth as a function
  of time. The time axis starts January 1 and covers a year. Solid curves in
  SLI halo model (we also include possible velocity dispersions); dashed curve
  in the standard halo model. The phase of the modulation is opposite in the
  two models.}
\end{figure}

Fig.~3 clearly shows that the modulation in Sikivie's late-infall halo model
has a phase opposite to the modulation in the standard halo model. The maximum
of the mean WIMP speed on Earth occurs in early June in the standard model but
in early December in Sikivie's late-infall halo model.  With our assumptions and
approximations, the maximum mean speed occurs June 1 in the standard model and
December 12 in the SLI model; the minimum occurs December 1 in the standard
model and June 11 in the SLI model. We remind the reader that the uncertainties
in the Sun peculiar velocity and the approximations in the velocity of the
Earth amount to an uncertainty of a few days in the position of the maxima and
minima.

The change of phase in the modulation is independent of the velocity dispersion
assumed in the SLI model.  A dispersion of 30 km/sec gives very similar results
to the zero dispersion (see Fig.~3). Even a dispersion of 220 km/s (not
plotted), similar in magnitude to the average velocity, would give a phase of
the modulation equal to the zero dispersion case (but the value of the average
relative velocity would be much larger).

The local dark halo density in Sikivie's late-infall halo model, obtained by
summing up the densities in each flow in Table 1, is $\rho$= 0.37 GeV/cm$^3$.
The upper bounds on the scattering cross section for a particular WIMP depend
on the value of the product $\rho \langle v(t) \rangle$.  The bounds obtained
with Sikivie's late-infall model and with a standard model with the same local
density would differ by the ratio of WIMP velocities in Fig.~3.  Otherwise the
ratio of the local densities in both models should be taken into account.

\section{Recoil-energy spectrum}

The WIMP detection rate per unit detector mass and nucleus recoil energy in the
range $(E, E+dE)$ can be written as
\begin{equation}
  R(E) = \frac{ \rho \sigma_0 F^2(q) } { m \mu^2 } \eta(E,t).
\end{equation}
Here $\sigma_0$ is a normalized WIMP-nucleus cross section; $\mu = m M / (m+M)$
is the reduced mass of the WIMP-nucleus system ($m$ is the WIMP mass and $M$ is
the nucleus mass); $F(q)$ is a nuclear form factor function of the nucleus
recoil momentum $q = 2 M E$; $\eta(E,t)$ determines the annual modulation and
depends only on the WIMP velocity distribution in the rest frame of the Earth
$f_{\oplus}({\bf u},t)$,
\begin{equation}
  \eta(E,t) = \frac{1}{2} \int d\Omega_u \,
  \int_{\sqrt{ME/2\mu^2}}^\infty u \, f_{\oplus}({\bf u},t) \, du.
\end{equation}

The nuclear form factor $F(q)$ depends on the type of WIMP-nucleus
interactions, namely if they are spin-dependent or spin-independent, and
reflects the mass and spin distributions inside the nucleus. It may strongly
affect the counting rates in dark matter detectors, but since it is
time-independent it is inessential in the analysis of the annual modulation.
Therefore in the examples below we set $F(q)=1$. The recoil-energy spectra we
plot should be multiplied by the appropriate $F(q)$.

We now focus on $\eta(E,t)$ since it is the only time dependent part in the
recoil-energy spectrum due to WIMP-nucleus collisions.  For the standard halo
model,
\begin{eqnarray}
  \eta_{\rm std}(E,t) & = & \frac{1}{N_{\rm esc}} \left\{ 
    \frac{1}{4 v_{\oplus}(t)} \left[ 
      {\rm erf}\!\left( \frac{ w+v_{\oplus}(t) }{ \overline{v}_0 } \right) - 
      {\rm erf}\!\left( \frac{ w-v_{\oplus}(t) }{ \overline{v}_0 } \right)  
    \right] \right. \nonumber \\ && \hspace{0.5in} \left. -
    \frac{1}{\pi^{1/2} \overline{v}_0} e^{-v_{\rm esc}^2/\overline{v}_0^2} 
  \right\}
\end{eqnarray}
with $w = \sqrt{ME/2\mu^2}$. In Sikivie's late-infall model without velocity
dispersion, each flow contributes a flat spectrum up to the maximum recoil
energy
\begin{equation}
  E_{{\rm max},i}(t) = \frac{ 2 \mu^2}{M} \big| 
   {\bf v}_i - {\bf v}_{\odot} - {\bf v}_{\oplus}(t) \big|^2 .
\end{equation}
The resulting spectrum is a series of steps where each step corresponds to a
flow,
\begin{equation}
  \eta_{\rm SLI}(E,t) = \sum_i \frac{\rho_i}{\rho} \, 
\frac{\theta(E_{{\rm max},i}(t) -
  E) }{2 | {\bf v}_i - {\bf v}_{\odot} - {\bf v}_{\oplus}(t) |} .
\end{equation}
Here $\theta$ is the Heaviside function. Notice that the position of the steps
given by $E_{{\rm max},i}(t)$ depends on time, and so the end-point of each
step is annually modulated. 

In Fig.~4 we plot the recoil-energy spectra at the maximum and minimum of the
modulation in the mean WIMP speed on Earth. At the lowest recoil-energies the
phase of the modulation is opposite in the two models; at intermediate energies
it is the same; at higher energies it is opposite again. To produce the figure
we have assumed a local halo density $\rho=0.3{\rm GeV}/c^2/{\rm cm}^3$, a WIMP
mass $m=60 {\rm GeV}/c^2$, a ${}^{73}$Ge detector, a WIMP-nucleus cross section
$\sigma_0 = 10^{-35} {\rm cm}^2$, and a nuclear form factor $F(q)=1$.
Corresponding spectra for other detectors can be obtained using the relations
$E' = (\mu'/\mu)^2 (M/M') E$ and $R' = F^2(q) (\sigma'_0/\sigma_0) (\mu/\mu')^2
R$.

One may worry that the lowest-energy step does not persist when additional
flows with smaller velocities are added in the SLI model. We argue that this
will not happen because while their velocities are smaller and smaller in the
galactic rest frame they are not smaller and smaller in the Sun or Earth rest
frames (see column 5 of Table 1).  Hence the end-point of the lowest-energy
step, which corresponds to the flow with the smallest speed relative to the
Earth, will not change if additional flows above the 20th are added in the
model.

Fig.~5 shows how the recoil-spectrum is modulated at fixed recoil-energies
$E=15,25,35$ keV.  The change of phase of the modulation is again visible. It
is also clear that except at the lowest recoil energies the spectral modulation
in the SLI halo model is no longer well approximated by a sinusoidal function.
This is due to the modulation of the position of the steps at $E_{{\rm
    max},i}(t)$ mentioned above. If particle flows are introduced in the halo
model, a more general analysis of the experimental data than that used up to
now may be called for.

\begin{figure}
\begin{center}
\includegraphics[width=\textwidth]{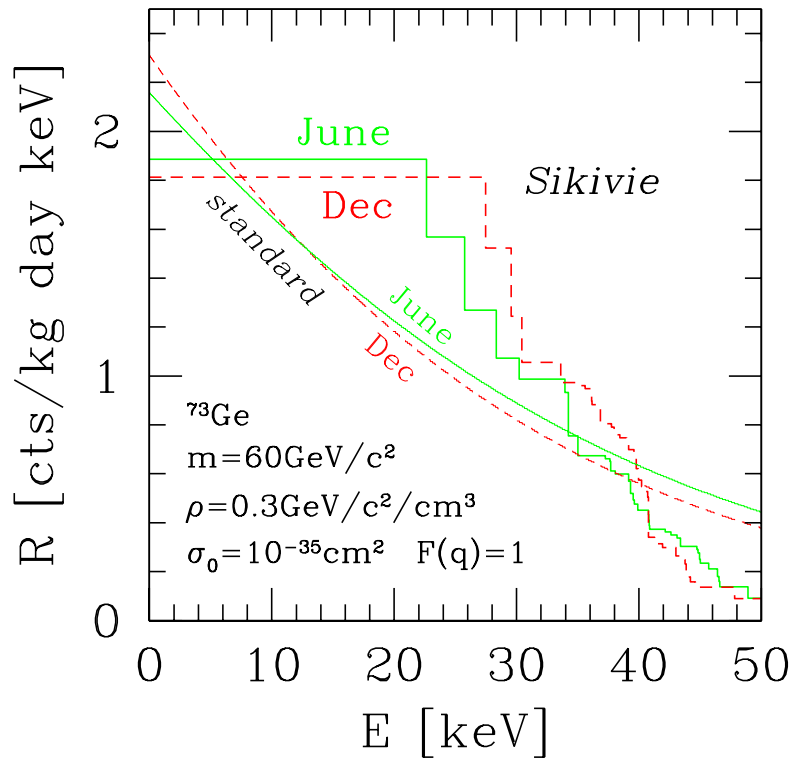}
\end{center}
\caption{Recoil-energy spectrum at the maximum and minimum of the 
  annual modulation in the mean WIMP speed on Earth.  Step-like curves in the
  Sikivie's late-infall halo model; smooth curves in the standard halo model.
  The phase of the modulation in the two models is opposite at low energies,
  the same at intermediate energies, opposite again at high energies. Notice
  that the end-point energy of each step in the SLI model is annually
  modulated.  }
\end{figure}
\begin{figure}
\begin{center}
\includegraphics[width=\textwidth]{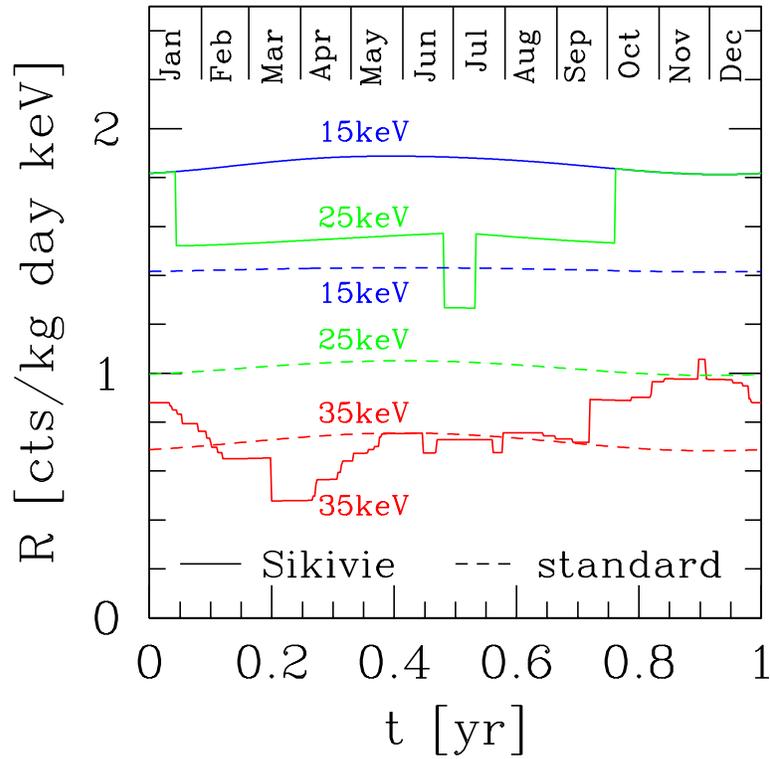}
\end{center}
\caption{Annual modulation of the recoil-energy spectrum at several fixed
  energies. Solid curves in the Sikivie's late-infall halo model; dashed curves
  in the standard halo model. Beyond the end-point energy of the lowest step,
  the modulation of the recoil-energy spectrum in the SLI halo model may be
  poorly approximated by a sinusoidal (cfr.\ the 25keV curve, e.g.). }
\end{figure}

The effect of a velocity dispersion in each flow is to blur the end-point of
each step. The recoil-spectrum in the SLI model with velocity dispersion can be
obtained using the expression of $\eta_{\rm std}(E,t)$ after taking the limit
$v_{\rm esc}\to\infty$ and replacing $1/N_{\rm esc} \to \sum_i \rho_i/\rho$ and
$v_{\oplus}(t) \to | {\bf v}_i - {\bf v}_{\odot} - {\bf v}_{\oplus}(t) |$.

In Sikivie's model there is no virialized halo component. If one should exist,
as assumed in Ref.~\cite{copi}, the local halo velocity distribution and local
WIMP fluxes would be a combination of both components. This would result in a
superposition with some relative amplitude of the fluxes and rates shown in
Figs.~1, 2, 4, and 5. The phase of the annual modulation would then depend on
the relative strength of both components.
 
\section{Conclusions}

The late-infall model~\cite{LImodel} of halo formation, modified in recent
years by P.~Sikivie and collaborators \cite{SI,STW,Slecture} to include a net
angular momentum and axial symmetry around its direction, predict
non-virialized flows of dark matter particles falling into the galaxy and
oscillating in and out many times. Hence it predicts a local velocity
distribution completely different from the standard truncated Gaussian usually
assumed.  P.~Sikivie in Ref.~\cite{Slecture} has given the local velocities and
densities of the first twenty pairs of flows (the first pair corresponding to
particles coming for the first time into the galaxy from opposite sides of it,
the second to those passing for the second time, etc.)  in his particular
late-infall model that has parameters that fit well the halo of our galaxy.  In
this model most WIMPs come from directions in the hemisphere opposite to the
Sun motion.  Namely the average ``WIMP wind'' velocity on Earth (as pointed out
by Sikivie himself \cite{Slecture}) is reversed with respect to that in the
standard halo model. Thus, in Sikivie's halo model, the annual modulation of a
galactic WIMP signal has a phase opposite to the usually assumed. This was
illustrated in Fig.~3. Moreover (as shown in Fig.~2) many WIMPs in this halo
model approach the Sun and the Earth from directions above and below the
galactic plane, with a clear pattern of directions. This would be very
important for experiments in which directionality could be measured.

The main message we want to convey is that if a non-virialized halo component
due to the infall of (collisionless) dark matter particles cannot be rejected,
an annual modulation in a dark matter signal should be looked for by
experimenters without fixing the phase a-priori.  The detection of an annually
modulated dark matter signal would not only reveal the nature of the dark
matter in our halo, but the structure of the dark matter halo itself.

\section*{Acknowledgements}

We thank the Aspen Center for Physics where this work was discussed with
P.~Sikivie last June. As we finished writing this paper, a similar work by A.M.
Green appeared \cite{green}: we basically agree with her conclusions where we
overlap.

\end{document}